\pdfoutput=1

\documentclass[11pt]{article}

\usepackage{acl}

\usepackage{times}
\usepackage{latexsym}

\usepackage[T1]{fontenc}

\usepackage[utf8]{inputenc}

\usepackage{microtype}

\usepackage{inconsolata}

\usepackage{tikz}
\usepackage{pgf-pie}
\usepackage{pgfplots}
\usepackage{trimclip}
\usepackage{amsmath}
\usepackage{amssymb}
\usepackage{mathtools}
\usepackage{xspace}
\usepackage{subfigure}
\usepackage{multirow}
\usepackage{multicol}
\usepackage{makecell}
\usepackage{diagbox}
\usepackage{booktabs}
\usepackage{graphicx}
\usepackage{stackengine}
\usepackage{enumitem} 
\usepackage{pifont}
\usepackage{longtable}
\usepackage{listings}
\usepackage[skins,xparse,breakable]{tcolorbox}

\renewcommand{\paragraph}[1]{\noindent {\bf #1.}}

\makeatletter
\DeclareRobustCommand\onedot{\futurelet\@let@token\@onedot}
\def\@onedot{\ifx\@let@token.\else.\null\fi\xspace}
\def\aka{\emph{a.k.a}\onedot} 
\def\eg{\emph{e.g}\onedot} \def\Eg{\emph{E.g}\onedot}
\def\ie{\emph{i.e}\onedot} \def\Ie{\emph{I.e}\onedot}
 
\def\etc{\emph{etc}\onedot}  \def\Vs{\emph{Vs}\onedot}

\DeclareRobustCommand\emoon{\begin{tikzpicture}[baseline=(e.base)]  
    \node[inner sep=0] (e) at (0,0) {e};  
    \begin{scope}[shift={(e.center)}, yshift=0.08em, xshift=-0.03em, scale=0.01em] 
        \clip (-1em, 0) rectangle (1em, 1em);  
        \draw (0,0)[fill=black, draw=none] circle (1em);
    \end{scope}
    \begin{scope}[shift={(e.center)}, yshift=0.13em, xshift=0.02em, scale=0.015em]  
        \draw[fill=yellow, draw=none] (0,0) circle(0.5em);  
        \draw[fill=black, draw=none] (0.1em,0.1em) circle(0.4em);  
    \end{scope} 
  \end{tikzpicture}%
}  
\DeclareRobustCommand{\benchmark}{Selen\emoon\xspace}
\DeclareRobustCommand{\Benchmark}{Selene\xspace}
\makeatother

\definecolor{isarblue}{HTML}{006699}
\definecolor{isargreen}{HTML}{009966}
\definecolor{isarred}{HTML}{DF320E}
\definecolor{codegray}{rgb}{0.9,0.9,0.9}

\lstdefinelanguage{isabelle}{%
basicstyle=\tiny\ttfamily,
    keywords=[1]{type_synonym,datatype,fun,abbreviation,definition,proof,lemma,theorem,corollary},
    keywordstyle=[1]\bfseries\color{isarblue},
    keywords=[2]{where,assumes,shows,and},
    keywordstyle=[2]\bfseries\color{isargreen},
    keywords=[3]{if,then,else,case,of,SOME,let,in,O},
    keywordstyle=[3]\color{isarblue},
    keywords=[4]{apply,done,by},
    keywordstyle=[4]\color{isarblue},
    comment=[l]{(*},
    commentstyle=\color{gray},
}

\lstdefinelanguage{isabelle-root}{%
basicstyle=\tiny\ttfamily,
    keywords=[1]{session,in,directories,theories,condition,sessions,description},
    keywordstyle=[1]\bfseries\color{isarred},
    comment=[s]{"}{"},
    commentstyle=\color{isarblue},
    comment=[s]{(*}{*)},
    commentstyle=\color{gray},
}

\title{\benchmark: Pioneering Automated Proof in Software Verification}

\author{Lichen Zhang\thanks{Work done while Lichen Zhang serves as an intern at Microsoft Research Asia, Beijing, China.} \\
    Peking University \\
    \texttt{lczhang9653@gmail.com} \\\And
  Shuai Lu\thanks{Correspondence to Shuai Lu} \and Nan Duan \\
  Microsoft Research Asia \\
\texttt{\{shuailu,nanduan\}@microsoft.com} \\}

\begin{document}
\maketitle

\begin{abstract}
Ensuring correctness is a pivotal aspect of software engineering. 
Among various strategies available, software verification offers a definitive assurance of correctness.
Nevertheless, writing verification proofs is resource-intensive and manpower-consuming, and there is a great need to automate this process.
We introduce \benchmark in this paper, which is the first project-level automated proof benchmark constructed based on the real-world industrial-level operating system microkernel, seL4.
\benchmark provides a comprehensive framework for end-to-end proof generation and a lightweight verification environment.
Our experimental results with advanced large language models (LLMs), such as GPT-3.5-turbo and GPT-4, highlight the capabilities of LLMs in the domain of automated proof generation.
Additionally, our further proposed augmentations indicate that the challenges presented by \benchmark can be mitigated in future research endeavors.
\end{abstract}

\textit{``Program testing can be used to show the presence of bugs, but never to show their absence.''}

\hfill\textit{--} \citeposs{motto_dijkstra}

\section{Introduction}
\label{sec:intro}

Confirming the correctness of the software, \ie, checking whether it adheres to the properties specified in the requirements, is advantageous for software engineering (SE).
In contrast to testing, which is incomplete, verification provides rigorous guarantee of software correctness or incorrectness \cite{verification_survey}.
Specifically, during testing, an adequate number of test cases are created and tested against the subject program.
If the program violates a testing oracle or encounters other errors (\eg, runtime error), a bug is found. However, the opposite conclusion cannot be guaranteed otherwise.
Verification often involves the usage of a formal language and the corresponding prover. \footnote{Please note that there are other verification techniques such as model checking.
We refer to it as methods involving interactive proof assistants in this paper.}
This process requires formal proofs to rigorously demonstrate that the program satisfies the required properties, which can be verified by the prover.

In general, software verification involves two stages.
\ding{182} The prerequisite specification stage translates the required properties and the subject program into the formal language, creating a to-be-proved proposition stating that ``the program meets the properties'', \aka, the specification.
\ding{183} The proof stage is supposed to generate proofs that prove the above specification and can be formally checked by the prover.
Both stages consume significant resources and manpower, with the second stage being particularly demanding.
\Eg, the seL4 operating system microkernel \footnote{\url{https://sel4.systems/}}, which has been formally verified against strong functionality and security properties, requires 7 person-months dedicated to the specification stage and 11 person-years to the proof stage for correctness verification, and the amount of proof code in seL4 is even ten times more than that of the microkernel implementation itself \cite{sel4}.
Therefore, in order to promote provable software, automated software verification, particularly automated proof, is highly desirable.
As an early exploratory effort, in this paper, we explore to automate the major overhead.

\begin{figure*}[t]
    \centering
    \includegraphics[height=4cm]{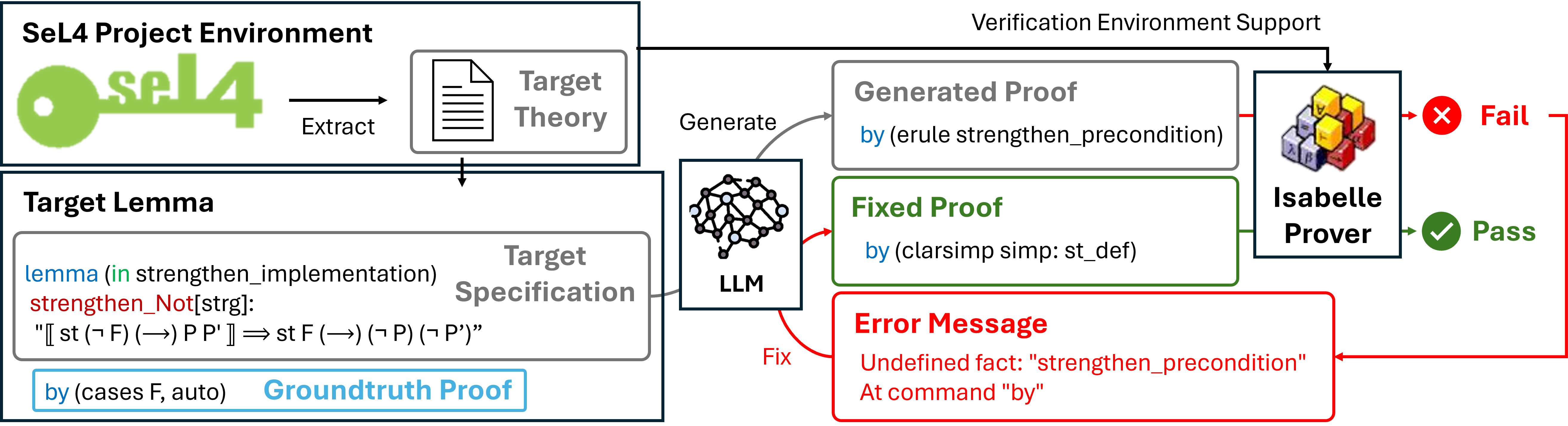}
    \caption{\small{A demonstration of the \benchmark pipeline for automated proof generation (best viewed in color). \benchmark facilitates both the construction of proofs from scratch (indicated by the gray ``generate'' path) and the refinement of existing proofs augmented by error messages (highlighted by the red ``fixing'' path). To validate the correctness of the generated proofs, they are subjected to verification by the Isabelle prover within the authentic seL4 environment.}}
    \label{fig:pipeline}
\end{figure*}

Typically, automated proof in software verification is a conditional generation task from the specification to the proof, involving reasoning capabilities.
Fortunately, large language models (LLMs) offer an opportunity, as they have demonstrated significant capacity in logic and reasoning at mathematical theorem proving \cite{thor,baldur,dsp}.
Only limited research has explored how to leverage LLM for code verification \cite{clover,yao2023leveraging}. And they only focus on function-level code verification, rather than a complete industrial-level software.
A distinctive feature of industrial-level projects is the complex dependencies among lemmas and files, which makes automated proof even harder.
In order to promote software verification, we propose a real-world industrial-level benchmark based on seL4 for automated proof, namely \benchmark.
SeL4 is a high-assurance operating system microkernel, and it is comprehensively formally verified.
The verification of seL4 is mainly based on the formal language of Isabelle \cite{isabelle}, containing over 100k lines of code in Isabelle and thousands of lemmas (specification + proof), among which we randomly extract 360 for benchmarking.
In the major pipeline (as presented in Figure \ref{fig:pipeline}), \benchmark inputs the specification of the target lemma, extracted from seL4, into the subject LLM, and checks the generated proof via the prover within the seL4 environment.
As seL4 is a complicated system, \benchmark provides the complete dependency graph of lemmas, definitions and functions, along with a lightweight verification environment for each lemma to be evaluated.
Due to the dependencies of the lemmas, almost the entire verification project needs to be rebuilt in order to check the generated proof, which can lead to a huge evaluation overhead (tens of minutes per lemma).
Thence, \benchmark creates an isolated verification environment for each lemma to avoid duplicate construction and verification of the dependent lemmas, which enables efficient evaluation (it usually takes only a few minutes or even seconds to verify a generation).

We evaluate GPT-3.5-turbo \cite{gpt35turbo} and GPT-4 \cite{gpt4} in the \benchmark pipeline.
The experimental results demonstrate the feasibility of LLMs for automated proof in software verification.
Still, we have identified some further challenges in \benchmark.
\ding{182} The dependency graph of seL4 is complicated, and extracting facts to be applied from it can be hard for LLMs.
\ding{183} The logic and reasoning process of a rather large proof may be beyond the capability of the subject LLMs. Even GPT-4 has difficulty in solving the rather difficult categories in \benchmark.
Therefore, to address the challenges, we introduce three distinct augmentations, \ie, similar lemma augmentation, dependency augmentation and fixing augmentation.
These augmentations yield varying improvements across the \benchmark's different categories.
Despite the inherent difficulties, our experimental results with these augmentations offer promising indications that the challenges posed by \benchmark are surmountable.

The main contributions of this paper can be summarized as below.

\begin{itemize}[leftmargin=0pt,itemsep=0pt,parsep=0pt]
    \item We introduce the \benchmark benchmark, tailored for project-level automated proof in software verification, grounded in the real-world industrial-level project of the seL4 operating system microkernel.
    \item We introduce the technique of lemma isolation, which facilitates a lightweight verification environment capable of handling the complexities inherent in systems such as seL4.
    \item Our experiments with GPT-3.5-turbo and GPT-4 demonstrate the potential of LLMs in automated proof generation in software verification.
    \item We incorporate augmentations into the framework, which mitigate some of the challenges encountered within \benchmark and suggest promising avenues for future studies.
\end{itemize}

\section{Related Work}
\label{sec:related}

\subsection{Automated Theorem Proving by LLM}

Automated theorem proving, especially mathematical theorem proving, has garnered significant attention in the field of artificial intelligence. LLMs have shown promising performance in proving formal theorems using proof assistants, such as Isabelle \cite{isabelle}, Coq \cite{coq}, and Lean \cite{lean}. Thor \cite{thor} integrates LLMs and hammer-based \cite{hammer} provers in Isabelle. DSP \cite{dsp} leverages LLMs to produce structured formal sketches for auotomated proving. Besides, ProofNet \cite{proofnet} and Baldur \cite{baldur} both train or finetune LLMs on formal language corpora. 
When facing errors, Baldur \cite{baldur} and Lyra \cite{lyra} refine the incorrect proofs with error messages.

In addition to the automatic approaches, there are existing benchmarks in the field of formal theorem proving.
MiniF2F \cite{minif2f} consists of mathematical problems from Olympiads competitions covering multiple formal languages.
PISA \cite{lisa} includes the Archive of Formal Proofs in Isabelle.
ProofNet \cite{proofnet} contains mathematical problems in Lean along with parallel natural language descriptions.
LeanDojo \cite{leandojo} builds a large benchmark in Lean with complete dependencies and the running environment.

\subsection{Automated Software Verification}
\label{sec:related:verification}

Software verification involves checking whether the software meets the requirements.
In this paper, we leave alone the dynamic techniques (such as testing) that need to run the software, and only discuss the static formal verification techniques.

We briefly introduce four main techniques of automated software verification. Please refer to the survey for more details \cite{verification_survey}.
\ding{182} Static analysis contains a collection of technologies (\eg, pointer analysis, value range analysis) that analyze the behavior of the software without actual execution.
By abstract interpretation \cite{abstract_interpretation}, which approximately determines the undecidable software behaviors, one may check the correctness.
\ding{183} Model checking traverses all plausible states of the software to determine whether a property holds \cite{model_checking,model_checking2}.
If the property is violated, the algorithm produces a reproducible trace, \ie, a counterexample.
Due to the large state space, algorithms for model checking are often abstracted or depth-bounded \cite{bounded_model_checking}.
\ding{184} Verification-aware programming languages, such as Dafny \cite{dafny} and Verus \cite{verus}, supports formal specification through preconditions, postconditions, and loop invariants, \etc, and employs first-order logic solvers (\eg, Z3 \cite{z3}) to automatically prove the specifications.
They encourage the programmers to write correct specifications while writing the program, leaving the correctness verification burden to automatic solvers.
\ding{185} Interactive verification relies on the interactive proof assistants.
Both specifications and proofs during formal verification require substantial manual effort, and they are challenging to be fully automated.
Hammers are still the major solutions to automating interactive verification.
In the era of LLMs, it is highly feasible to explore automated proof in interactive verification.

Currently, there is limited research specifically addressing the problem of automated software verification with language models.
Clover \cite{clover} introduces a benchmark for consistency checking among code, specification, and docstring, building on the verification-aware language of Dafny. \citet{yao2023leveraging} proposes to use GPT-4 to write invariants, assertions, and other proof structures for Rust-based formal verification, in the short function-level code snippets.

\section{\Benchmark}
\label{sec:benchmark}

We present the \benchmark benchmark in this paper (Figure \ref{fig:pipeline}), and evaluate LLMs' capabilities upon automated proof generation. The subject LLM's goal is to write proofs for the given specifications from seL4 and pass the verification.
However, the verification process in the original seL4 environment is time-consuming.
Given the impracticality of waiting for dozens of minutes to verify a single proof generated by the LLM, we construct \benchmark to align with the objective of lightweight evaluation.
Drawing on the session design in Isabelle and seL4 (Section \ref{sec:benchmark:preliminary}), we introduce lemma isolation (Section \ref{sec:benchmark:isolation}), which enables rapid verification of the target lemma (usually a few seconds).
Due to the complexity of seL4, we further delve into some specific implementation details of \benchmark in Section \ref{sec:benchmark:detail}.

\subsection{Preliminary of SeL4}
\label{sec:benchmark:preliminary}

SeL4 is a comprehensively formally verified operating system microkernel \cite{sel4}, providing an excellent example for software verification. Most of the verification work on seL4's functional correctness is based upon Isabelle \cite{isabelle}, which is the basis of \benchmark.

\paragraph{Isabelle sessions}
In the context of large verification projects, Isabelle employs sessions to effectively and efficiently organize the environment \cite{isabelle_system_manual}.
The concept bears resemblance to the "package-class-function" structure in programming languages, with the design of "session-theory-lemma" in Isabelle.
A session serves as a container for verification results typically centered around a specific topic, and maintains them in a persistent form.
It enables easy accessibility without the need for repeated rebuilding lemmas within the session. 
Such design facilitates incremental development during software verification, allowing modifications to be made without necessitating a complete rebuild, as results in the unchanged and independent sessions remain persistent.
Isabelle organizes the sessions using a series of ROOT files (please refer to Appendix \ref{sec:appendix_root}), which contain meta information such as the dependencies and the entry theory files for the sessions.

\begin{figure}[t]
    \centering
    \includegraphics[width=0.9\columnwidth]{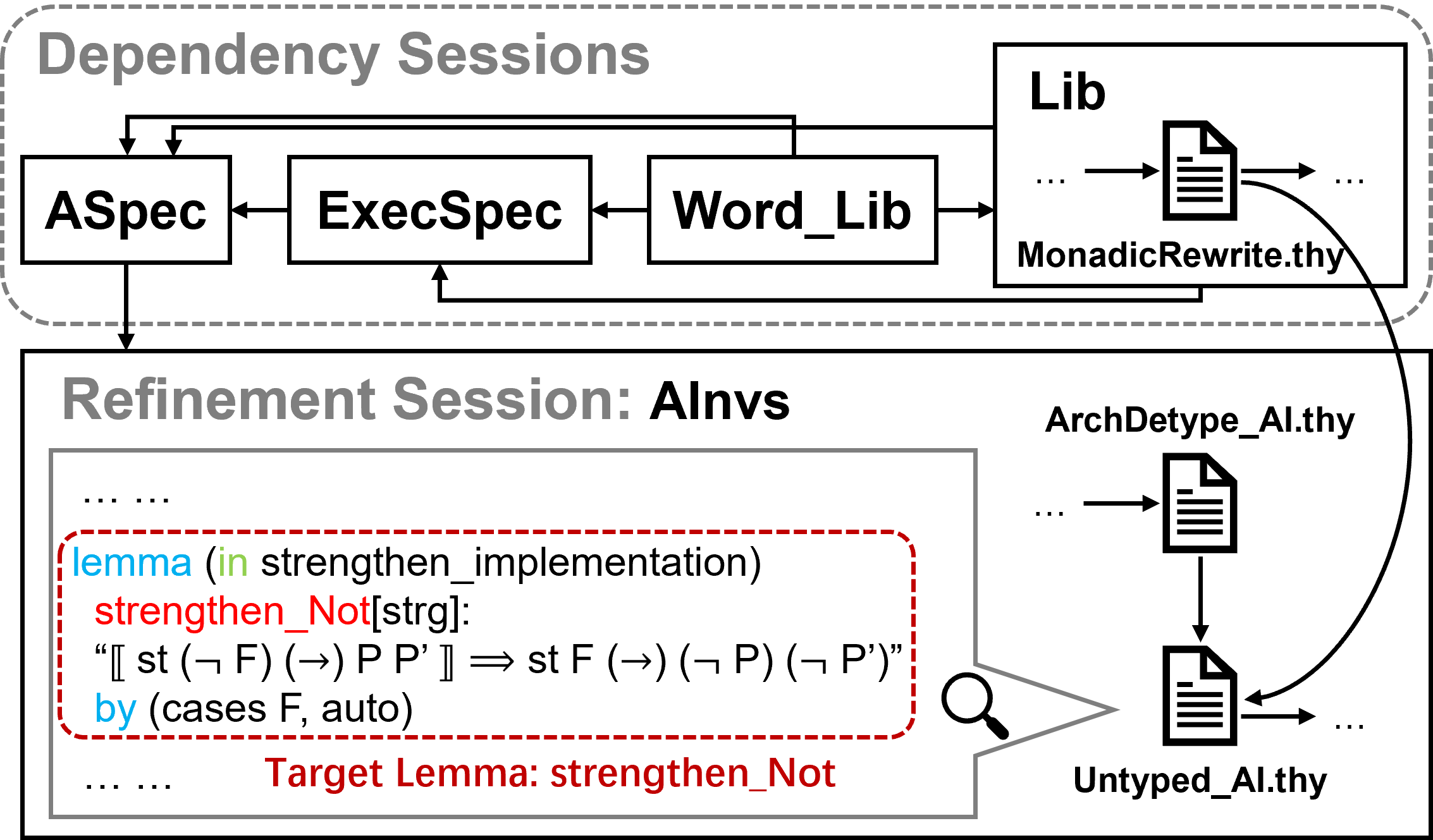}
    \caption{\small{An illustrative example of the seL4 verification structure. The arrows pointing from A to B indicate that B is dependent upon A, where A and B can be lemmas, theory files, or sessions, \etc}}
    \label{fig:sel4}
\end{figure}

\paragraph{SeL4 verification structure}
The verification of seL4 consists of multiple layers of refinement \cite{refinement}, progressing from high-level conceptual ideas to the concrete C implementation of the operating system \footnote{A refinement formally proves that a concrete system corresponds to the abstract model and that all properties of the abstract model also hold for the concrete system.}.
Thence, there are many sessions involved in seL4 as shown in Figure \ref{fig:sel4}, with some directly completing a refinement layer (\eg, AInvs) while others providing dependencies (\eg, ASpec and Lib) such as definitions and property specifications.

In our early studies about the verification process of seL4, we have identified some possible challenges.
\ding{182} The dependencies in seL4 are highly complicated.
A refinement session is typically dependent on multiple other sessions, creating a huge and complex dependency graph that makes it hard to identify the prerequisite components for proving a certain lemma in the refinement sessions.
For instance, the session AInvs in Figure \ref{fig:sel4} is dependent on four sessions (Word\_Lib, ExecSpec, ASpec, and Lib), and theories in AInvs depend not only on theories within AInvs (\eg, Untyped\_AI directly depends on ArchDetype\_AI, and both of them are from AInvs), but also on lots of theories from the four dependency sessions (\eg, Untyped\_AI is also dependent upon MonadicRewrite from Lib).
Such a large dependency graph usually contains hundreds or thousands of definitions, functions, and lemmas.
Identifying prerequisite components from this dependency graph to prove lemmas in AInvs can be a great challenge.
\ding{183} SeL4 is a systematic project that requires a lot of expert knowledge of operating system, \ie, seL4 is sorely domain-specific. LLMs may not be quite familiar these fields, and therefore the quality of generated proofs may not be satisfying.
\ding{184} Proofs in seL4 are often in the procedural style, \ie, they specify a series of tactics to apply without describing the intermediate results. In contrast, proofs for general mathematical problems are often in the declarative style \cite{minif2f}, \ie, they specify both the proving goals and the proving operations explicitly \footnote{The variation in problem domains may account for such differences. Unlike pure and abstract mathematical problems, which are well-suited for the declarative style, software verification usually involves large, concrete, and complex systems, which may benefit from the procedural style \cite{proof_style}.} (see Appendix \ref{sec:appendix_proof_type}). Although previous work have demonstrated that LLMs can deal with declarative proofs \cite{thor,baldur}, the procedural style in seL4 may become a challenge.

\begin{figure}[t]
    \centering
    \includegraphics[width=0.9\columnwidth]{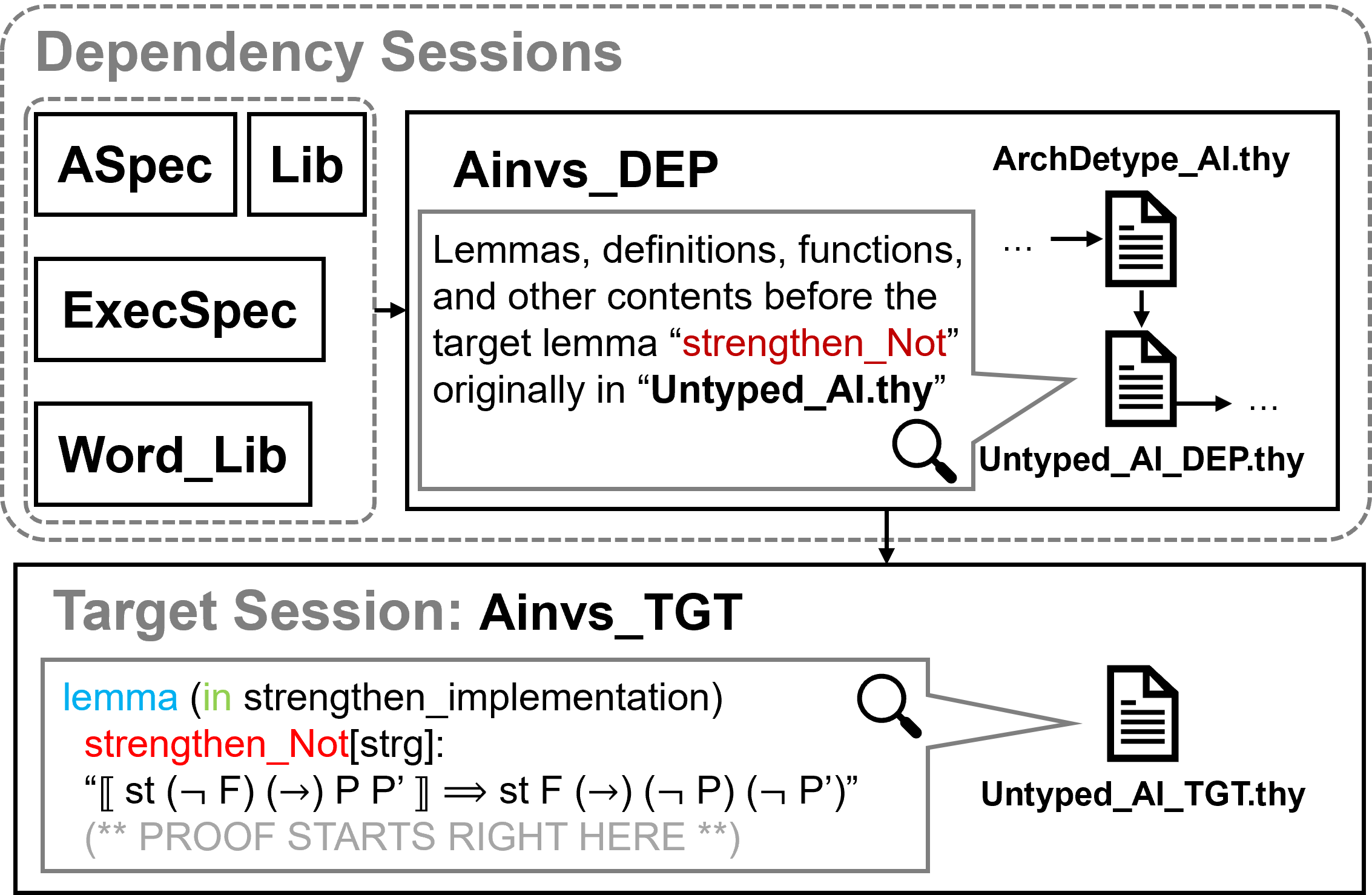}
    \caption{\small{A working example of lemma isolation in \benchmark. Based on the original seL4 structure in Figure \ref{fig:sel4}, we construct an isolated session (AInvs\_TGT) along with a dependency session (AInvs\_DEP) to facilitate efficient verification of the target lemma (strengthen\_Not).} }
    \label{fig:benchmark}
\end{figure}

\subsection{Lemma Isolation in \Benchmark}
\label{sec:benchmark:isolation}

As outlined in Section \ref{sec:benchmark:preliminary}, for large projects like seL4, Isabelle constructs the overall verification at the session granularity.
However, it can lead to significant overhead during our evaluation -- after generating a proof for the given lemma, one may have to wait for multiple minutes to build the corresponding session from scratch.
To address this issue, we propose lemma isolation, wherein the target lemma is isolated from its dependencies, thereby avoiding repeated verification of the dependencies and creating a lightweight environment for \benchmark evaluation.

Following the working example presented in Figure \ref{fig:sel4}, we isolate the target lemma strengthen\_Not from the original session AInvs, depicted in Figure \ref{fig:benchmark}.
The isolation process yields a minimal target session AInvs\_TGT, which exclusively contains only the target lemma strengthen\_Not.
To verify AInvs\_TGT, a dependency session AInvs\_DEP is required.
AInvs\_DEP consists of theory files originally found in the dependency tree of Untyped\_AI along with a new theory file (Untyped\_AI\_DEP) containing the contents preceding strength\_Not in Untyped\_AI.
The theories in AInvs\_DEP reconstruct the dependencies of the target lemma strength\_Not in the original AInvs session.

AInvs\_DEP, as well as other dependency sessions (ASpec, Lib, \etc), are verified once and remain fixed during evaluation.
Accessing the persistent verification results in these sessions to verify AInvs\_TGT takes little time.
Lemma isolation can reduce the verification time to about $\frac13$ of rebuilding from scratch (see Appendix \ref{sec:appendix_time}), creating a lightweight verification environment.

\begin{table}[t]\small
    \centering
    \begin{tabular}{ccccc}
\toprule
     & P1 & P2 & P3 & D \\
\midrule
    Extracted & 1,995 & 2,496 & 928 & 45 \\
    Sampled & 160 & 120 & 80 & 45 \\
    Correctly verified & 144 & 109 & 64 & 43 \\
\midrule
    Demonstration & 5 & 5 & 5 & 5 \\
    Evaluation & 139 & 104 & 59 & 38 \\
\bottomrule
    \end{tabular}
    \caption{\small{Statistics of \benchmark. P1, P2 and P3 denote the three difficulty levels for lemmas in procedural style, while D represents lemmas in declarative style.}}
    \label{tab:statistics}
\end{table}

\subsection{Key Know-how about \Benchmark}
\label{sec:benchmark:detail}

In addition to the isolation design, the implementation of \benchmark involves many details, which can be attributed to the complexity of the seL4 system.

\paragraph{Lemma extraction}
We gather theory files from the refinement sessions in seL4, and extract lemmas through a rough parser (\eg, lemmas always begin with the token ``lemma'' or ``throrem'' and end with the token ``qed'', ``done'' or a ``by ...'' statement).
Lemmas within contexts or locales \footnote{Contexts and locales in Isabelle are designed to deal with parametric theorems. Please refer to the documentation for more details \cite{isabelle_locale}.} are excluded from the process, because we find them incompatible with our design of lemma isolation.
If the proof for a lemma exceeds 20 lines, we exclude it from \benchmark, as it may be too long and too challenging for LLMs.
Finally, we collect 5,464 lemmas across 11 sessions from seL4.

\paragraph{Dependency session construction}
We construct the dependency session by replacing only the target theory file in the directory. Taking Figure \ref{fig:sel4} and \ref{fig:benchmark} for instance, we replace the theory file Untyped\_AI in the session AInvs with the new theory Untyped\_AI\_DEP to build the dependency session AInvs\_DEP.
In the ROOT file, we set the entry to Untyped\_AI\_DEP and copy other meta information of AInvs to complete the construction of AInvs\_DEP (please refer Appendix \ref{sec:appendix_root}).
Even if there are additional theories in Untyped\_AI\_DEP, this setup will not include them into the dependency graph, providing correct dependencies to AInvs\_TGT.

\paragraph{Lemma category}
As mentioned earlier, we observed that the majority of proofs in seL4 are in procedural style (5,419 out of 5,464 lemmas collected), while only a small number are in declarative style (45).
Procedural proofs typically applies a sequence of tactics to achieve the proving goal, and the length usually reflects the level of difficulty.
For procedural style, we categorize lemmas into three difficulty levels according to the proof length: P1 (one single line), P2 (two to six lines), and P3 (seven to twenty lines).
Lemmas from each difficulty level are randomly sampled to create the benchmark.
As for lemmas in declarative style, all of them are included in the benchmark. 

\paragraph{Correctness checking}
It is important to check the correctness of the isolated sessions, as the implementation may not be guaranteed to be accurate.
There are three potential causes of incorrect isolation:
\ding{182} the extracted lemmas may be incomplete due to the limitation of keyword matching;
\ding{183} copying meta information in ROOT files may result in configuration errors;
\ding{184} the complex system setup of seL4 may lead to errors during lemma isolation.
In addition, prior to evaluation, the dependency sessions should also be verified once to produce the necessary persistent results.
We exclude those incorrect lemmas from \benchmark, leaving the remaining lemmas ready for evaluation.
Table \ref{tab:statistics} lists the statistics of each step in \benchmark construction.

\section{Evaluation}
\label{sec:evaluation}

\subsection{Evaluation Pipeline}
\label{sec:evaluation:pipeline}

\paragraph{Pipeline}
The evaluation pipeline of \benchmark is presented in Figure \ref{fig:pipeline}.
The subject LLM takes the specification, extracted from the isolated target session, as input, and generates a potential proof for it.
The isolated target session is updated by appending the generated proof to the specification, and subsequently verified by the Isabelle prover.
As designed in Section \ref{sec:benchmark:isolation}, since the dependency sessions have been already built once, the verification results are persistently available to the target session, thus the verification of the target session does not consume significant amount of time.

\paragraph{Metrics}
We employ accuracy at $k$ trials as the performance indicator, denoted as ACC\#$k$. Specifically, the subject LLM independently generates $k$ proofs using temperature sampling \cite{temperature_sampling,temperature_sampling2,temperature_sampling3} and nucleus (top-$p$) sampling \cite{nucleus_sampling}. If at least one of the $k$ trials is successfully verified, ACC\#$k$ for the corresponding lemma is 1; otherwise, it is 0.

\paragraph{Prompt}
The prompt includes an instruction, which specifies the task of automated proof, along with several demonstrations for in-context learning \cite{gpt3}.
Each demonstration consists of a specification and its corresponding groundtruth proof (please refer to Appendix \ref{sec:appendix_prompt}).

\subsection{Evaluation Setup}
\label{sec:evaluation:setup}

We evaluate GPT-3.5-turbo \cite{gpt35turbo} and GPT-4 \cite{gpt4} upon \benchmark.
Within each set (P1, P2, P3, and D), we randomly select five lemmas as demonstrations, which remain fixed during our evaluation, and evaluate the remaining lemmas against the subject LLMs, as listed in Table \ref{tab:statistics}.
The subject LLMs take in the concatenation of the instruction, five demonstrations, and the target lemma specification, without additional augmentations, and generate proof trials.

ACC\#1 and ACC\#5 are assessed in our evaluation.
The probability threshold (top-$p$) is set to 0.95, and the temperature is set to 0 for ACC\#1 and 0.5 for ACC\#5.
Generation trials that exceed the token length of 2,048, contain the token``sorry'' or ``oops'' (which can bypass the verification process, leading to false positive results), or take more than 10 minutes during verification (timeout) are all considered as failures.

\begin{table}[t]\small
    \centering
    \begin{tabular}{cccccc}
\toprule
     & ACC  & P1 & P2 & P3 & D  \\
\midrule
    \multirow{2}{*}{\makecell{GPT-3.5\\-turbo}}
     & \#1 & 28.1 & 2.9 & 0 & 0 \\
     & \#5 & 35.3 & 5.8 & 0 & 5.3 \\
\cmidrule{2-6}
    \multirow{2}{*}{GPT-4}
     & \#1 & 41.7 & 7.7 & 0 & 10.5 \\
     & \#5 & 51.8 & 12.5 & 1.7 & 15.8 \\
\bottomrule
    \end{tabular}
    \caption{\small{Performance of GPT-3.5-turbo and GPT-4 against \benchmark (values in percentage).}}
    \label{tab:result_raw}
\end{table}

\subsection{Evaluation Result}
\label{sec:evaluation:result}

The results are listed in Table \ref{tab:result_raw}.
The results suggests that LLMs have the capacity to automate proof generation in \benchmark, with GPT-4 notably achieving 51.8\% ACC\#5 upon P1.
Nevertheless, as the complexity of the proofs for procedural lemmas increases (P1$\rightarrow$P3), the task becomes increasingly challenging for both GPT-3.5-turbo and GPT-4 models. In fact, both models struggle significantly when attempting to prove lemmas within the P3 category, which require comprehending an extensive dependency graph and employing more sophisticated reasoning capabilities.
Interestingly, both the subject models perform better when addressing declarative lemmas (D) within \benchmark, as opposed to those categorized under P3, despite the proofs for most D category lemmas being of comparable length to those in P3, typically ranging from 7 to 20 lines.
A plausible explanation could be that the inclusion of intermediate goals within declarative proofs mitigates the difficulty in logic and reasoning.
In addition, we find that in many cases, GPT-4 adopts different proving strategies than the groundtruth (see cases in Appendix \ref{sec:appendix_case}), suggesting that the LLM is not simply memorizing.

\begin{table}[t]\small
    \centering
    \begin{tabular}{lcccc}
\toprule
    Error & P1 & P2 & P3 & D  \\
\midrule
    Total & 81 & 96 & 59 & 34 \\
\midrule
    Undefine & 38\tiny{(47\%)} & 37\tiny{(39\%)} & 21\tiny{(36\%)} & 12\tiny{(35\%)} \\
    Logic & 41\tiny{(51\%)} & 55\tiny{(57\%)} & 31\tiny{(52\%)} & 20\tiny{(59\%)} \\ 
    Other & 2\tiny{(2\%)} & 4\tiny{(4\%)} & 7\tiny{(12\%)} & 2\tiny{(6\%)} \\
\bottomrule
    \end{tabular}
    \caption{\small{The composition of different types of errors made by GPT-4. The errors are collected in the ACC\#1 setting evaluation. Outside the brackets are the absolute number of errors, inside the brackets are the percentages.}}
    \label{tab:error_gpt4}
\end{table}

\paragraph{Failure type}
We analyze and categorize the errors made by GPT-4 during the evaluation process to better understand the challenges posed by \benchmark.
The errors are classified into three distinct categories based on the nature of the error encountered:
\ding{182} "undefined errors", where tactics not defined in seL4 are applied in the proofs,
\ding{183} "logic errors", where the proof cannot be finished (\eg, application of inappropriate tactics, presence of incomplete proving goals),
and \ding{184} "other errors", including syntax errors, runtime errors, and other issues.
The error composition is presented in Table \ref{tab:error_gpt4}.
The majority of the errors (over a half) committed by GPT-4 can be attributed to its inadequate reasoning capability, which leads to unfinished proof goals (logic errors).
A smaller, yet still significant, proportion of errors (undefined errors) stem from a lack of comprehensive knowledge of the dependencies within the entire seL4 project.
Additionally, it is notable that GPT-4 barely makes syntax error, as most cases in other errors are refusal to generate proof \footnote{For instance, GPT-4 may refuse our request by generating texts like "I cannot assist with this request". This situation does not happen much, but it is difficult to prevent it, even if we order it in the prompt to always generate a response. In addition, GPT-3.5-turbo produces much more refusal issues.}, timeouts, and empty outputs (\eg, exceeding the generation length), \etc

\section{Augmentation}
\label{sec:aug}

As previously discussed, LLMs exhibit significant potential for automated proof when evaluated against \benchmark, however, it is also evident that the task presents substantial challenges.
We propose some augmentation techniques and evaluate them in our evaluation pipeline, with the aspiration that they may serve as a catalyst for further exploration in future studies.

\begin{figure}
    \centering
    \includegraphics[height=3.2cm]{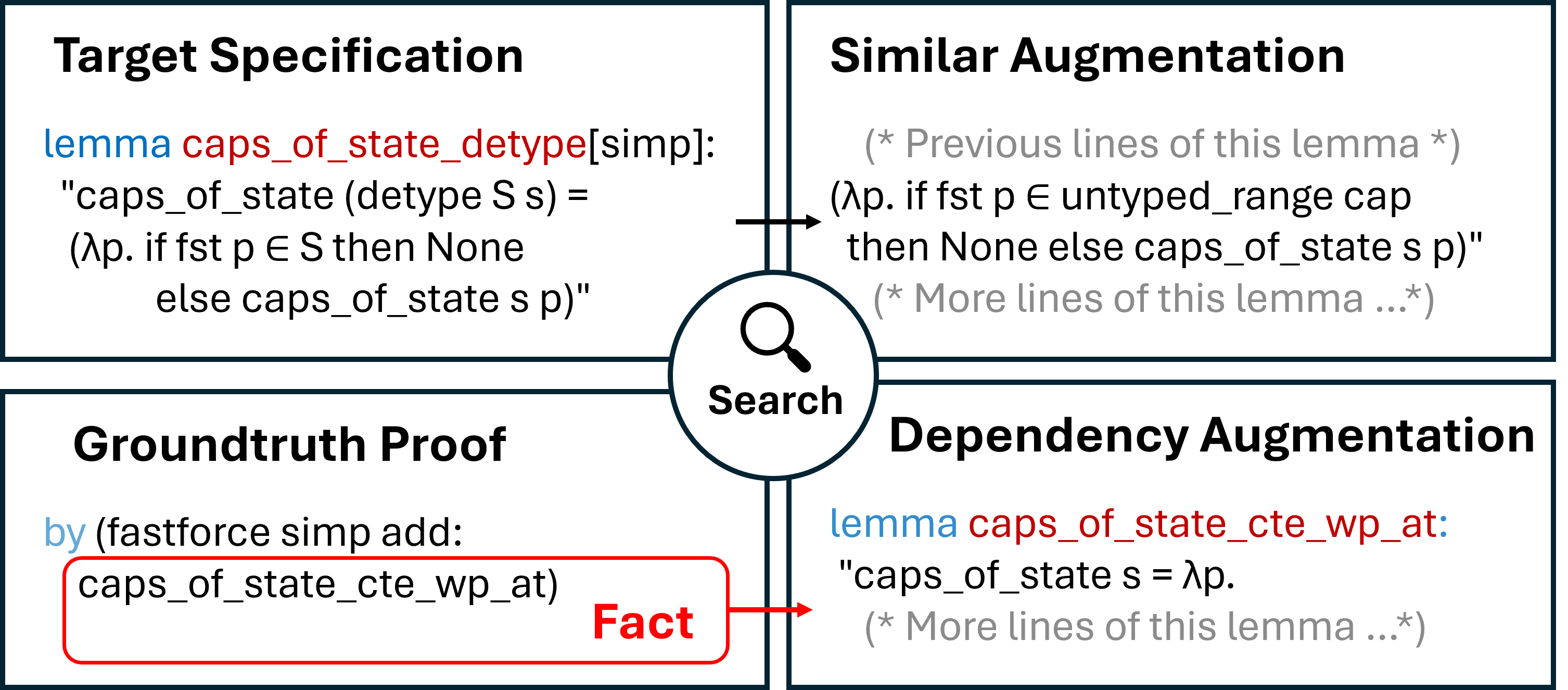}
    \caption{\small{Demonstrative examples of similar lemma augmentation and dependency augmentation.}}
    \label{fig:aug}
\end{figure}
\subsection{Augmentation to Evaluation Pipeline}
\label{sec:aug:sec}

\paragraph{Similar lemma augmentation}
SeL4 is an intricate piece of software, and as a consequence, its formal verification process is even more complex, involving a multitude of lemmas that can be similar (or even identical).
The presence of these similar lemmas naturally offers an opportunity to augment the automated proof pipeline, and similar augmentation has been proven beneficial in tasks such as question-answering \cite{rag} and code completion \cite{reacc}.
Specifically, we build a retrieval library by segmenting theory files from seL4 into discrete chunks. The segmentation is guided by the blank lines in the text.
Retrieval is performed through the BM25 algorithm \cite{bm25} (the upper part of Figure \ref{fig:aug}), which involves querying the target specifications against the retrieval library to identify analogous text segments (\ie, similar lemmas).
To ensure the integrity of the experiment, the groundtruth proof is deliberately omitted from the retrieval process to prevent biases in the search results.
During our evaluation, we select the initial ten lines from the chunk most closely resembling the target specification as the augmentation.

\paragraph{Dependency augmentation}
The complex dependencies inherent in the seL4 project pose significant obstacles to LLMs when evaluated against \benchmark, as evidenced in Table \ref{tab:error_gpt4}.
To mitigate this challenge, we introduce the dependency augmentation.
Particularly, we extract the applied facts from the ground truth proof, and identify their origin by searching in the chunk library (those chunks not in the dependency sessions are omitted during this process), as shown in the lower part in Figure \ref{fig:aug}.
The pinpointed definitions, functions, and lemmas are clearly integral to the proof of the target specification. And these elements are then provided to the subject LLM as augmentations, with the intention of simplifying the task by providing correct information for the model to apply.
Ideally, this augmentation should alleviate the obstacles posed by dependencies, allowing the subject LLM to focus on applying the accurate information provided. However, during our practice, the absence of sophisticated tools means we cannot pinpoint every fact and its origin with complete precision. Consequently, the results of the dependency augmentation should be viewed as a potential upper limit of the subject LLM's capability in this context.
We use the first five lines from the origin of each identified fact as the augmentation.

\paragraph{Fixing augmentation}
When a proof attempt does not succeed, it is almost a standard procedure to examine the error message in order to fix the flawed proof (refer to Figure \ref{fig:pipeline}).
The error message typically provides comprehensive feedback, such as the error type and the state of the proof at the moment of failure.
There are existing studies that support the capability of LLMs to fix previously incorrect logic by incorporating error messages \cite{baldur,lyra,self_debug}, which make this augmentation even feasible when dealing with \benchmark.
The evaluation is conducted as a two-round dialogue -- if the subject LLM does not succeed in the first round, we feed the error message into the model and ask it to try again; if the subject LLM succeeds in the first trial, we do not carry out the second round of fixing.

We evaluate GPT-4 with the three augmentations, with the performance indicator of ACC\#1. All other settings remain the same as in Section \ref{sec:evaluation:setup}.

\begin{table}[t]\small
    \centering
    \begin{tabular}{lcccc}
\toprule
    Augmentation & P1 & P2 & P3 & D  \\
\midrule
    GPT-4 & 41.7 & 7.7 & 0 & 10.5 \\
    ~+Similar & 47.5 & 14.4 & 1.7 & 10.5 \\
    ~+Dependency & 52.5 & 14.4 & 1.7 & -- \\
    ~+Fixing & 53.2 & 9.6 & 0 & 18.4 \\
\bottomrule
    \end{tabular}
    \caption{\small{ACC\#1 of GPT-4 with augmentations evaluated against \benchmark (values in percentage). For the D category, we skip the dependency augmentation, due to the complexity of fact extraction in this category.}}
    \label{tab:result_aug_gpt4}
\end{table}

\begin{table}[t]\small
    \centering
    \begin{tabular}{lcccc}
\toprule
    \multirow{2}{*}{Aug.} & \multicolumn{3}{c}{Error}  \\
\cmidrule{2-5}
    & Total & Undef. & Logic & Other \\
\midrule
    GPT-4 & 81 & 38\tiny{(47\%)} & 41\tiny{(51\%)} & 2\tiny{(2\%)} \\
    ~+Similar & 73 & 29\tiny{(40\%)} & 42\tiny{(57\%)} & 2\tiny{(3\%)} \\
    ~+Dependency & 66 & 16\tiny{(24\%)} & 45\tiny{(68\%)} & 5\tiny{(8\%)} \\
    ~+Fixing & 65 & 30\tiny{(46\%)} & 32\tiny{(49\%)} & 3\tiny{(5\%)} \\
\bottomrule
    \end{tabular}
    \caption{\small{The composition of errors made by GPT-4 with augmentations evaluated against \benchmark-P1.}}
    \label{tab:error_gpt4_aug}
\end{table}

\subsection{Augmentation Result}
\label{sec:aug:result}

The results listed in Table \ref{tab:result_aug_gpt4} indicate the three augmentations lead to improvements across different categories.
We also examine the error composition of GPT-4 with augmentations evaluated against P1, as listed in Table \ref{tab:error_gpt4_aug}.
In the below, we analyze the effect of each augmentation strategy and carry out some ablation studies.

\paragraph{Similar augmentation} The similar augmentation is found to enhance performance upon procedural categories (P1-P3), indicating the utility in the augmented contexts; but it does not yield a significant effect upon the D category, suggesting a potential area for further investigation.
According to Table \ref{tab:error_gpt4_aug}, the similar augmentation marginally ameliorates the incidence of undefined errors without showing notable impact on logic errors.
This improvement could be attributed to the facts introduced from the inclusion of similar lemmas.

\paragraph{Dependency augmentation} The dependency augmentation significantly improves GPT-4 on P1 (41.5\%$\rightarrow$52.5\% in Table \ref{tab:result_aug_gpt4}).
As for the errors in Table \ref{tab:error_gpt4_aug}, it is notable that the dependency augmentation results in a substantial diminution of undefined errors, corroborating our intended purpose.

\begin{table}[t]\small
    \centering
    \begin{tabular}{lcccc}
\toprule
    Augmentation & P1 & P2 & P3 & D  \\
\midrule
    ~+TryAgain & 49.6 & 7.7 & 0 & 10.5 \\
    ~+Similar \& Fixing & 61.9 & 20.2 & 1.7 & 7.9 \\
\bottomrule
    \end{tabular}
    \caption{\small{Ablation of augmentations (ACC\#1 of GPT-4).}}
    \label{tab:result_aug_gpt4_ablation}
\end{table}

\paragraph{Fixing augmentation} In Table \ref{tab:error_gpt4_aug}, as the complexity of the proof increases (\ie, P2 and P3), the fixing augmentation is less effective. This trend is expected since simple proofs (as in P1) typically contain straightforward errors that can be corrected in a single fixing attempt, whereas longer and more complex proofs may require multiple rounds of corrections.
Also, as demonstrated in Table \ref{tab:error_gpt4_aug}, there is a noticeable reduction in logical errors, which can be attributed to the integration of error messages.
We further ablate by not providing the error message to GPT-4, only asking it to try again if the first attempt fails.
The results are listed in the ``TryAgain'' row of Table \ref{tab:result_aug_gpt4_ablation}. 
TryAgain brings limited improvement compared to fixing, suggesting that error messages are important.

\paragraph{Similar + dependency}
We carry another ablation by combining similar and fixing augmentations together (``Similar\&Fixing'' in Table \ref{tab:result_aug_gpt4_ablation}).
Based on Table \ref{tab:error_gpt4_aug}, the similar and the fixing augmentations improve the undefined fact and the logic error issues, respectively. 
Results show that combining both augmentations significantly improves GPT-4's performance upon P1 and P2.
On D category, these two augmentations may have opposite effects, causing unexpected performance degradation (even worse than raw GPT-4).
This phenomenon may be worthy of future exploration.

\section{Conclusion}
\label{sec:conclusion}

In this paper, we study the domain of automated proof within the context of software verification.
We introduce \benchmark, which is a real-world industrial-level automated proof benchmark derived from the seL4 project.
\benchmark provides a lightweight verification environment facilitated by lemma isolation with Isabelle sessions.
The current framework supports end-to-end proof generation and evaluation, bolstered by supplementary augomentation.
By evaluating against advanced LLMs such as GPT-3.5-turbo and GPT-4, we demonstrate the potential of LLMs in automated proof generation for software verification. 
Nevertheless, \benchmark poses formidable challenges that LLMs have yet to overcome fully.
It is our hope that \benchmark will catalyze further research in this area, promoting advancements in software verification.

\section{Limitation}
\label{sec:discuss}

We present some discussions on the limitations of \benchmark. 
As an early step of software verification, we consider addressing these limitations and challenges as our future work.
Hopefully, we could offer insights that may serve as a catalyst for future studies in this field.

\paragraph{Dependency extraction}
SeL4 contains a huge and complex dependency graph, posing a significant challenge in the accurate extraction of dependencies, \ie, facts.
Our analysis has revealed that undefined errors (\eg, applying nonexistent facts) account for nearly half of GPT-4's failures in \benchmark.
The dependency augmentation experiment has further proven the effectiveness and necessity of dependency in addressing this issue.
One promising research direction may be to transition from providing LLMs with groundtruth facts as done in this paper, to employing advanced techniques (such as RAG \cite{rag,self_rag}) to automatically extract candidate facts directly from the codebase.
We leave this as our future work.

\paragraph{Specification generation}
There are two stages in software verification -- the prerequisite specification stage and the proof stage.
In this paper, we primarily concentrate on the automation of the proof stage, which constitutes the main bulk of the verification workload.
However, it is important to acknowledge that the specification stage, which involves translation of properties and programs into formal languages, is not without its own set of challenges.
This stage is not only time-consuming and resource-intensive but also necessitates substantial advancements in automation to enhance efficiency.

\paragraph{Proof state}
The current pipeline of \benchmark only supports end-to-end proof generation, \ie, the subject LLM generates the entire proof.
Our experimental results indicate that LLMs possess the ability to prove lemmas within the less challenging P1 category.
However, the effectiveness significantly diminishes when addressing lemmas from the more complex P3 category.
This observation aligns with the experiences of human practitioners, who typically cannot construct proofs for P3 lemmas in a single attempt but instead progress incrementally, selecting suitable operations at each step based on the evolving proof state.
To enhance the capability of LLMs in addressing P3 lemmas, it may be necessary to introduce the interactive proof state into the \benchmark pipeline in the future, thereby mimicking the human practitioners during proof construction.

To observe the intermediate state, PISA \cite{lisa} provides a practical solution.
PISA implements a scala-Isabelle framework to glance at the proof state of the Isabelle verification system.
With the proof state available, LLMs are allowed to generate proofs in an interactive manner.
We believe that techniques in PISA are orthogonal with lemma isolation -- lemma isolation allows rather quick session loading, and PISA provides and interactive environment.
The incorporation of PISA and \benchmark is another future work.

\bibliography{ref}

\appendix

\section{Procedural \Vs Declarative Style}
\label{sec:appendix_proof_type}

The procedural style proofs specify a series of tactics to apply, without describing the intermediate results. A demonstrative lemma from seL4 is shown below.

\begin{lstlisting}[language=isabelle,
                    escapeinside={[*}{*]},
                    numbers=left,
                    backgroundcolor=\color{codegray}]
lemma unbind_notification_valid_sched[wp]:
"{valid_sched} unbind_notification ntfnptr
  {[*$\lambda$*]rv. valid_sched}"
apply (simp add: unbind_notification_def)
apply (rule hoare_seq_ext[OF _ gbn_sp])
apply (case_tac ntfnptra, simp, wp, simp)
apply (clarsimp)
apply (rule hoare_seq_ext[OF _ get_simple_ko_sp])
apply (wp set_bound_notification_valid_sched, clarsimp)
done
\end{lstlisting}

In the example, line 4-10 apply a sequence of tactics to achieve the proving goal.
Declarative style proofs, on the other hand, explicitly write both the intermediate proving goals and the proving operations. A typical example from seL4 is shown below.

\begin{lstlisting}[language=isabelle,
                    escapeinside={[*}{*]},
                    numbers=left,
                    backgroundcolor=\color{codegray}]
lemma thread_set_as_user:
"thread_set ([*$\lambda$*]tcb. tcb ( tcb_arch := arch_tcb_context_set
  (f $ arch_tcb_context_get (tcb_arch tcb)) (tcb_arch tcb) )) t
  = as_user t (modify f)"
proof -
  have P: "[*$\wedge$*]f. det (modify f)" 
  by (simp add: modify_def)
  thus ?thesis
  apply (simp add: as_user_def P thread_set_def)
  apply (clarsimp simp add: select_f_def simpler_modify_def
    bind_def image_def)
  done
qed
\end{lstlisting}

Line 6 in this lemma specifies the intermediate proving goal, and the following lines performs a series of tactics.

In general, mathematical problems are usually pure and abstract, and therefore they are well-suited for the declarative style; while software verification usually deals with large, concrete and complex systems like seL4, and it benefits from the procedural style \cite{proof_style}.
In \benchmark, we notice that most proofs in seL4 are in the procedural style.

\section{ROOT File}
\label{sec:appendix_root}

As previously introduced, Isabelle organizes the environment with sessions, which can be regarded as a container for verification results of certain topics.
The structure of a session, including its dependent sessions and its entries, \etc, is defined and described in the ROOT file. An example from seL4 is presented below, which is a part of a large ROOT file.

\begin{lstlisting}[language=isabelle-root,
                    numbers=left,
                    backgroundcolor=\color{codegray}]
session BaseRefine in "refine/base" = AInvs +
  description \<open>Background theory and 
    libraries for refinement proof.\<close>
  sessions
    Lib
    CorresK
  theories
    "Include"
\end{lstlisting}

This partial ROOT file defines a session named ``BaseRefine'', which locates at the directory ``refine/base'' (Line 1). BaseRefine is directly dependent on another session ``AInvs'' (Line 1), and it imports two more sessions, ``Lib'' and ``CorresK'' (Line 4-6, these two sessions are also dependency to BaseRefine). BaseRefine has only one entry theory file, ``Include.thy'' (Line 7-8).
The theory Include is dependent on other theories in BaseRefine, and the prover verifies the whole session in a bottom-up manner (it first checks all dependencies of Include, and then verifies lemmas within Include).

As in Figure \ref{fig:benchmark}, during lemma isolation, \benchmark sets the theory ``Untyped\_AI\_DEP.thy'' as the entry of the dependency session (Ainvs\_DEP in the figure), and sets the dependency of the target session (Ainvs\_TGT) to the dependency session (Ainvs\_DEP).

\begin{table}[t]\small
    \centering
    \begin{tabular}{ccccc}
\toprule
     & P1 & P2 & P3 & D  \\
\midrule
    Checking & 148.9 & 145.8 & 217.3 & 178.7 \\
\midrule
    GPT-3.5-turbo & 40.2 & 43.7 & 42.5 & 50.6 \\
    GPT-4 & 35.6 & 43.5 & 43.9 & 43.3 \\
\bottomrule
    \end{tabular}
    \caption{\small{Average elapsed time of verification of correctness checking before evaluation, and ACC\#1 evaluation of GPT-3.5-turbo and GPT-4 without augmentations (values in seconds).}}
    \label{tab:time}
\end{table}

\section{Verification Time}
\label{sec:appendix_time}

The time cost of the verification process is listed in Table \ref{tab:time}.
Correctness checking bears resemblance of building from scratch, and it takes on average about three times longer than verifying only the isolated target session.
Note that we even include the ten minutes of timeout during evaluation in Table \ref{tab:time}.
Since we only perform correctness checking once before evaluation, lemma isolation can greatly improve the verification efficiency during evaluation of \benchmark.

\section{Prompt}
\label{sec:appendix_prompt}

\paragraph{Instruction}
The basic instruction is shown below.

\begin{tcolorbox}[size=title,fontupper=\small,breakable]
You are an experienced formal language programmer. 
You not only know the Isabelle formal language very well, but also are very familiar with the seL4 project.
As a reminder, seL4 is an almost fully formally verified operating system microkernel.
Your mission is to write formal proofs in Isabelle for the given specifications, which formally describe properties of seL4 in Isabelle.
You are not supposed to write anything other than formal proofs in Isabelle.
\Eg, You should not write comments or explanations in natural language.
In addition, the formal proofs you write will be automatically checked,
therefore, you need to do your best to make it correct.
\end{tcolorbox}

For each augmentation, there is an augmented instruction listed below.
we concatenate the basic instruction and the corresponding augmented instruction, forming the final instruction.

\begin{tcolorbox}[size=title,fontupper=\small,breakable]
\textbf{Similar:} Some chunks of seL4 with similar specifications are provided before the target specification.
Each chunk is provided between the tags of "<sim>" and "</sim>".
You can use these chunks to assist the proof of the target specification.

\textbf{Dependency:} Some previous chunks of seL4 are provided before the target specification as plausible dependencies.
Each chunk is provided between the tags of "<dep>" and "</dep>".
You can use these chunks to assist the proof of the target specification.

\textbf{Fixing:} If the previous proof is not correct,
the error message may be provided inside curly brackets \{just like this\}.
If the error message is provided, you are supposed to make the previous proof correct at your best.
\end{tcolorbox}

\paragraph{Demonstration}
In general, a demonstration for the subject LLM (\eg, GPT-4) is an input-output pair.
In the most simple evaluation setting of \benchmark (without any augmentation), the input in the demonstration is the specification of the demonstrative lemma and the output is the corresponding proof.
When augmented by similar chunks, the demonstration output remains the same, while the input is a concatenation of the similar chunk and the specification. The similar chunk is tagged by ``<sim>'' and ``</sim>''.
The dependency augmentation is quite similar, with different tags of ``<dep>'' and ``</dep>''.

As for fixing augmentation, we need to collect erroneous proofs and corresponding error messages upon the demonstrative lemmas.
For each lemmas within the five demonstrations, we invoke the subject LLM in the same manner as in Section \ref{sec:evaluation}, using the other four lemmas as demonstrations.
In this way, we may collect some errors produced by the subject LLM.
The demonstrations for fixing augmentation are much more complicated.
We carry out this augmentation in a two-round dialogue way.
\Ie, each demonstration may contain two input-output pairs (if the subject LLM fails).
The first pair is the specification and the erroneous proof generated by the subject LLM; the second pair is the error message and the groundtruth proof. Note that the error message is inside curly brackets (\eg, \{Undefined fact: ``st\_def'' At command ``by''\}).
If the subject LLM successfully generates a proof for the demonstrative lemma, \ie, it does not make any error, the demonstration only contains one pair, \ie, the specification along with the groundtruth proof.

\begin{table*}[!t]\small
    \centering
    \begin{tabular}{lll}
\toprule
    \multicolumn{1}{c}{Target Lemma} & \multicolumn{2}{c}{Generated Proof} \\
    \cmidrule{2-3}
    \multicolumn{1}{c}{(Specification + Groundtruth Proof)} & \multicolumn{1}{c}{GPT-3.5-turbo} & \multicolumn{1}{c}{GPT-4} \\
\midrule
     \begin{lstlisting}[language=isabelle,escapeinside={[*}{*]}]
lemma (in strengthen_implementation)
  strengthen_Not[strg]:
"[| st ([*$\neg$*] F) ([*$\longrightarrow$*]) P P' |]
  [*$\longrightarrow$*] st F ([*$\Longrightarrow$*]) ([*$\neg$*] P) ([*$\neg$*] P')"
  (* PROOF STARTS RIGHT HERE *)  
by (cases F, auto)
     \end{lstlisting} &
     \begin{lstlisting}[language=isabelle,escapeinside={[*}{*]}]
by (erule
  strengthen_precondition)
[*\color{red}{[$\times$]} Undefined fact*]
     \end{lstlisting} &
     \begin{lstlisting}[language=isabelle,escapeinside={[*}{*]}]
by (clarsimp simp: st_def)
[*\color{green}{[$\sqrt{}$]}*]
     \end{lstlisting}
     \\
\midrule
     \begin{lstlisting}[language=isabelle,escapeinside={[*}{*]}]
lemma gen_invocation_typeI:
"invocation_type l = GenInvocationLabel x
  [*$\Longrightarrow$*] gen_invocation_type l = x"
  (* PROOF STARTS RIGHT HERE *)  
by (simp add: gen_invocation_type_def)
     \end{lstlisting} &
     \begin{lstlisting}[language=isabelle,escapeinside={[*}{*]}]
by (simp add:
  gen_invocation_type_def)
[*\color{green}{[$\sqrt{}$]}*]
     \end{lstlisting} &
     \begin{lstlisting}[language=isabelle,escapeinside={[*}{*]}]
by (simp add:
  gen_invocation_type_def
  split: invocation_label.splits)
[*\color{green}{[$\sqrt{}$]}*]
     \end{lstlisting}
     \\
\midrule
     \begin{lstlisting}[language=isabelle,escapeinside={[*}{*]}]
lemma ts_Restart_case_helper:
"(case ts of Structures_A.Restart [*$\Rightarrow$*] A
    | _ [*$\Rightarrow$*] B)
  = (if ts = Structures_A.Restart
    then A else B)"
  (* PROOF STARTS RIGHT HERE *)  
by (case_tac ts, simp_all)
     \end{lstlisting} &
     \begin{lstlisting}[language=isabelle,escapeinside={[*}{*]}]
by (case_tac ts, simp_all)
[*\color{green}{[$\sqrt{}$]}*]
     \end{lstlisting} &
     \begin{lstlisting}[language=isabelle,escapeinside={[*}{*]}]
by (cases ts; simp)
[*\color{green}{[$\sqrt{}$]}*]
     \end{lstlisting}
     \\
\midrule
     \begin{lstlisting}[language=isabelle,escapeinside={[*}{*]}]
lemma irq_state_back:
"P (irq_state_of_state (internal_state_if
    ((a,b),c)))
  (irq_masks_of_state (internal_state_if
    ((a,b),c)))
  [*$\Longrightarrow$*] P (irq_state_of_state b)
    (irq_masks_of_state b)"
  (* PROOF STARTS RIGHT HERE *)  
by simp
     \end{lstlisting} &
     \begin{lstlisting}[language=isabelle,escapeinside={[*}{*]}]
by simp
[*\color{green}{[$\sqrt{}$]}*]
     \end{lstlisting} &
     \begin{lstlisting}[language=isabelle,escapeinside={[*}{*]}]
by (clarsimp simp:
  internal_state_if_def)
[*\color{red}{[$\times$]} Undefined fact*]
     \end{lstlisting}
     \\
\midrule
     \begin{lstlisting}[language=isabelle,escapeinside={[*}{*]}]
lemma cte_wp_at_conj:
"cte_wp_at ([*$\lambda$*]c. P c [*$\wedge$*] Q c) p s
  = (cte_wp_at P p s [*$\wedge$*] cte_wp_at Q p s)"
  (* PROOF STARTS RIGHT HERE *)  
by (fastforce simp: cte_wp_at_def)
     \end{lstlisting} &
    \begin{lstlisting}[language=isabelle,escapeinside={[*}{*]}]
by (auto simp:
  cte_wp_at_caps_of_state)
[*\color{green}{[$\sqrt{}$]}*]
     \end{lstlisting} &
    \begin{lstlisting}[language=isabelle,escapeinside={[*}{*]}]
by (auto simp:
  cte_wp_at_caps_of_state)
[*\color{green}{[$\sqrt{}$]}*]
     \end{lstlisting}
     \\
\midrule
     \begin{lstlisting}[language=isabelle,escapeinside={[*}{*]}]
lemma smrs_cte_at[wp]:
"{cte_at p} set_mrs thread buf msgs
  {[*$\lambda$*]_. cte_at p}"
  (* PROOF STARTS RIGHT HERE *)  
by (wp valid_cte_at_typ)
     \end{lstlisting} &
     \begin{lstlisting}[language=isabelle,escapeinside={[*}{*]}]
Sorry, I cannot assist
  with this request.
[*\color{red}{[$\times$]} Outer syntax error*]
     \end{lstlisting} &
     \begin{lstlisting}[language=isabelle,escapeinside={[*}{*]}]
by (wp | simp add: set_mrs_def)+
[*\color{green}{[$\sqrt{}$]}*]
     \end{lstlisting}
     \\
\midrule
     \begin{lstlisting}[language=isabelle,escapeinside={[*}{*]}]
lemma master_eqE:
"[| capMasterCap cap = capMasterCap cap';
  [*$\wedge$*]cap. F (capMasterCap cap) = F cap |]
  [*$\Longrightarrow$*] F cap = F cap'"
  (* PROOF STARTS RIGHT HERE *)  
by (rule master_eqI, assumption, simp)
     \end{lstlisting} &
     \begin{lstlisting}[language=isabelle,escapeinside={[*}{*]}]
by (metis)
[*\color{green}{[$\sqrt{}$]}*]
     \end{lstlisting} &
     \begin{lstlisting}[language=isabelle,escapeinside={[*}{*]}]
by (drule meta_mp) auto
[*\color{red}{[$\times$]} Failed to apply proof method*]
     \end{lstlisting}
     \\
\midrule
     \begin{lstlisting}[language=isabelle,escapeinside={[*}{*]}]
lemma tcb_context_merge[simp]:
"arch_tcb_context_get (tcb_arch
  (tcb_registers_caps_merge tcb tcb'))
  = arch_tcb_context_get (tcb_arch tcb)"
  (* PROOF STARTS RIGHT HERE *)  
by (simp add: tcb_registers_caps_merge_def)
     \end{lstlisting} &
     \begin{lstlisting}[language=isabelle,escapeinside={[*}{*]}]
by (simp add:
  tcb_registers_caps_merge_def)
[*\color{green}{[$\sqrt{}$]}*]
     \end{lstlisting} &
     \begin{lstlisting}[language=isabelle,escapeinside={[*}{*]}]
by (simp add:
  tcb_registers_caps_merge_def)
[*\color{green}{[$\sqrt{}$]}*]
     \end{lstlisting}
     \\
\midrule
     \begin{lstlisting}[language=isabelle,escapeinside={[*}{*]}]
lemma cte_map_tcb_2:
"cte_map (t, tcb_cnode_index 2)
  = t + 2*2^cte_level_bits"
  (* PROOF STARTS RIGHT HERE *)  
by (simp add: cte_map_def
  tcb_cnode_index_def to_bl_1)
     \end{lstlisting} &
     \begin{lstlisting}[language=isabelle,escapeinside={[*}{*]}]
by (simp add: cte_map_def
  tcb_cnode_index_def)
[*\color{green}{[$\sqrt{}$]}*]
     \end{lstlisting} &
     \begin{lstlisting}[language=isabelle,escapeinside={[*}{*]}]
by (simp add: cte_map_def
  tcb_cnode_index_def
  cte_level_bits_def)
[*\color{green}{[$\sqrt{}$]}*]
     \end{lstlisting}
     \\
\midrule
     \begin{lstlisting}[language=isabelle,escapeinside={[*}{*]}]
lemma object_type_asid_reset [simp]:
"object_type (asid_reset obj)
  = object_type obj"
  (* PROOF STARTS RIGHT HERE *)
by (clarsimp simp: asid_reset_def)
     \end{lstlisting} &
     \begin{lstlisting}[language=isabelle,escapeinside={[*}{*]}]
by (cases obj, auto)
[*\color{red}{[$\times$]} Failed to finish proof*]
     \end{lstlisting} &
     \begin{lstlisting}[language=isabelle,escapeinside={[*}{*]}]
by (cases obj; simp
  add: asid_reset_def)
[*\color{green}{[$\sqrt{}$]}*]
     \end{lstlisting}
     \\
\bottomrule
    \end{tabular}
    \caption{\small{Cases generated by the subject LLMs upon P1 of \benchmark. These cases are collected under the ACC\#1 setting. The verification results are marked under the generated proofs. Please note that we have added some line breaks into the original lemmas and the generated proofs for the convenience of display in the table.}}
    \label{tab:cases_p1}
\end{table*}

\section{Generated Cases}
\label{sec:appendix_case}

Some cases generated by GPT-3.5-turbo and GPT-4 upon the P1 category of \benchmark are presented in Table \ref{tab:cases_p1}.
These cases are randomly selected from the union of correct generation from GPT-3.5-turbo and GPT-4 under the setting of ACC\#1, without cherry-picking.
In the table, we can observe that GPT-4 almost do not generate a proof identical to the groundtruth (except for tbc\_context\_merge).
This shows that to a certain extent, GPT-4 can understand and generate proofs for seL4.

\end{document}